**Negative flat band magnetism in a spin–orbit-coupled correlated kagome magnet**

**Authors**
Jia-Xin Yin[1]*, Songtian S. Zhang[1]*, Guoqing Chang[1]*, Qi Wang[2], Stepan Tsirkin[3], Zurab Guguchia[1,4], Biao Lian[5], Huibin Zhou[6,7], Kun Jiang[8], Ilya Belopolski[1], Nana Shumiya[1], Daniel Multer[1], Maksim Litskevich[1], Tyler A. Cochran[1], Hsin Lin[9], Ziqiang Wang[8], Titus Neupert[3], Shuang Jia[6,7], Hechang Lei[2] & M. Zahid Hasan[1,10]†

**Affiliations**
[1]Laboratory for Topological Quantum Matter and Advanced Spectroscopy (B7), Department of Physics, Princeton University, Princeton, NJ, USA.
[2]Department of Physics and Beijing Key Laboratory of Opto-electronic Functional Materials & Micro-nano Devices, Renmin University of China, Beijing, China.
[3]Department of Physics, University of Zurich, Winterthurerstrasse 190, Zurich, Switzerland.
[4]Laboratory for Muon Spin Spectroscopy, Paul Scherrer Institute, CH-5232 Villigen PSI, Switzerland.
[5]Princeton Center for Theoretical Science, Princeton University, Princeton, NJ, USA.
[6]International Center for Quantum Materials and School of Physics, Peking University, Beijing, China.
[7]CAS Center for Excellence in Topological Quantum Computation, University of Chinese Academy of Science, Beijing, China.
[8]Department of Physics, Boston College, Chestnut Hill, MA, USA.
[9]Institute of Physics, Academia Sinica, Taipei, Taiwan.
[10]Lawrence Berkeley National Laboratory, Berkeley, CA, USA.
*These authors contributed equally.

†e-mail: mzhasan@princeton.edu

**It has long been speculated that electronic flatband systems can be a fertile ground for hosting novel emergent phenomena including unconventional magnetism and superconductivity[1-14]. Here we use scanning tunnelling microscopy to elucidate the atomically resolved electronic states and their magnetic response in the kagome magnet[15-20] $Co_3Sn_2S_2$. We observe a pronounced peak at the Fermi level, which is identified to arise from the kinetically frustrated kagome flatband. Increasing magnetic field up to ±8T, this state exhibits an anomalous magnetization-polarized Zeeman shift, dominated by an orbital moment in opposite to the field direction. Such negative magnetism can be understood as spin-orbit coupling induced quantum phase effects[21-25] tied to non-trivial flatband systems. We image the flatband peak, resolve the associated negative magnetism, and provide its connection to the Berry curvature field, showing that $Co_3Sn_2S_2$ is a rare example of kagome magnet where the low energy physics can be dominated by the spin-orbit coupled flatband. Our methodology of probing band-resolved ordering phenomena such as spin-orbit magnetism can also be applied in future experiments to elucidate other exotic phenomena including flatband superconductivity and anomalous quantum transport.**

Exploring the interplay between symmetry-breaking order and electronic topology in a strong coupling setting[1-4] is emerging as the new frontier in fundamental physics, which not only helps develop new concepts on phases of matter, but also provides indispensable knowledge to achieve future applications of quantum materials. The spin-orbit coupling (SOC) effects in flatband systems have recently been highlighted as extraordinary pursuits as they can embrace strong interaction and non-trivial topology with taking advantage of both[1-14]. Flatbands are rare and emerge only in a few systems such as twisted bilayer graphene, kagome lattices, and heavy fermion compounds. They can have two different origins: either they arise due to the lattice geometry as in the former two examples, or due to strongly localized atomic orbitals, as in the latter example. Benefiting from the quenched kinetic energy, flatbands can host correlated electronic states including ferromagnetism, superconductivity and Wigner crystallization[8-11]. In the presence of SOC and time-reversal symmetry breaking, flatbands can further acquire a non-zero Chern number arising from a nontrivial Berry phase, with the ability to support interacting topological phases including the fractional quantum Hall state[11-14]. Despite these exciting theoretical prospects, experimental progress remains haltingly elusive, due to the scarcity of real materials with the flatband at the Fermi level and the challenges in microscopically demonstrating the emergent behavior in such systems.

Here we research the atomically resolved electronic structure of kagome magnet[15-20] $Co_3Sn_2S_2$ at 4.2K under a tunable magnetic field with high-resolution scanning tunnelling microscopy/spectroscopy (STM/S). We observe a pronounced peak at the Fermi level associated with the momentum integrated state of the kagome flatband and discover a magnetization-polarized Zeeman effect. We theoretically study the relationship between kagome flatband, SOC effect, magnetism and the Berry curvature field on the magnetic ground state of this compound. All our results and theoretical analysis taken together uncover that the dominant contribution of the orbital magnetism of the flatband originates from the Berry curvature field in $Co_3Sn_2S_2$, which in this particular case leads to a band-specific strong diamagnetism and other anomalous behavior. Our work does show that $Co_3Sn_2S_2$ is a rare example of kagome magnet where the low energy physics can be dominated by the spin-orbit flatband.

$Co_3Sn_2S_2$ has a layered crystal structure with space group $R\bar{3}m$ and hexagonal lattice constants $a$ = 5.3 Å and $c$ = 13.2 Å (Fig. 1**a**). It consists of a kagome $Co_3Sn$ layer sandwiched between two hexagonal S layers, which are further sandwiched between two hexagonal Sn layers. The material has a ferromagnetic ground state (Curie temperature, $T_C$ = 170K) with the magnetization arising mainly from Co and aligned along the *c* axis[15,16]. The large electron negativity difference leads to the strongest bonds forming between the Co and S atoms (Fig. 1**b**). Upon cleaving, the S-Sn bond is broken yielding the S surface and the Sn surface (Fig. 1**c**). However, as these two layers have the same lattice symmetry, it is challenging to distinguish them by STM imaging.

Experimentally, we find that only 5% of the cleaving surfaces are atomically flat. We observe atomic surface steps consistent with the expected height of $c/3$ (Fig. 1**d**), and we find two kinds of surfaces with the same hexagonal symmetry and lattice constant around $a$: one has a large corrugation with adatom-defects (Fig. 1**e**), while the other has a smaller corrugation with vacancy-defects (Fig. 1**f**). As Sn has a larger atomic radius than S, the Sn surface can exhibit a smaller atomic corrugation (Fig. 1**c**). Moreover, since the Sn and S layers have a subatomic distance in the bulk, their bonding competes with the in-plane Sn-Sn bond. This competition can cause incomplete cleaving which creates Sn vacancies on the Sn surface and leaves Sn adatoms on the S surface. Based on their surface corrugations and atomic defects, we can identify these two surfaces as marked in Fig. 1**e** and **f**. Further conclusive evidence can be found by resolving the boundary between these two surfaces (Fig. 1**c**). An extensive search in over 30 samples systematically allowed us to observe a few cases where these two surfaces meet with an atomically resolved structure. Corroborating the previous identification, a representative topographic image (Fig. 2**a**) clearly demonstrates the coverage of the Sn adatoms gradually increasing on the S surface, eventually forming an atomically flat Sn surface.

Having identified these two surfaces at the atomic scale, we characterize their electronic structure by measuring the differential conductance on the clean areas (Fig. 2**b**). We find that the spectrum on the S surface exhibits a sharp peak at $E_P$ = -8meV with more states below $E_F$ (Fermi energy), whereas the Sn surface has more states above $E_F$ and no detectable peak. These spectral features are reproducibly observed on different samples. Notably, the sharp peak does not shift with increasing temperature up to 45K (Fig. 2**b** inset), nor is there any enhanced scattering vector at this energy when we analyze the spectroscopic map over a large S surface by Fourier transform (Fig. 2**d**). These observations essentially exclude a simple charge or magnetic order origin of this state (for example, as a coherent peak of an asymmetric energy gap), which often exhibits a strong temperature dependent evolution in energy as well as a well-defined wave vector in the quasi-particle scattering channel. Moreover, as the sharp peak is present only on one surface, whether it is a surface state or a projection of the bulk state needs be determined.

To address this question experimentally, we designed a clamp sample holder to allow us to perform side cleaving and tunnel to the side surface with atomic precision, which would produce a surface environment different from the above two surfaces. Such an experiment is challenging and sharp atomically resolved side cleaved surfaces for layered materials have been rarely reported due to the lack of a natural cleavage plane. Through multiple trials on a large number of samples, we were able to obtain a surface with quasi-atomic lattice structure (Fig. 2**e**) and detectable Bragg peaks in its Fourier transform image (Fig. 2**e** inset). Notably, we observe a broader peak on this surface (Fig. 2**c**), located at the same energy as the sharp peak observed on the S surface despite the large difference in their surface environments. This observation indicates that this state is unlikely to be a simple surface effect and is more likely to be an intrinsic band structure feature.

To examine the band structure origin of this peak, we perform first-principles calculations on its electronic structure. The calculated orbital resolved local density of states (LDOS) for the Sn-terminated surface, the S-terminated surface and the bulk are shown in Fig. 3**a**, respectively. In all cases, the Co 3*d* orbital is always the strongest, which agrees with the general consensus that the low energy physics of transition metal based materials are often dominated by the 3*d* electrons, and that the related tunnelling spectra are mainly associated with the 3*d* orbital[26]. For the two surface cases, there exists a noticeable peak at low energy only for the S-terminated surface as shown in the upper and middle panels of Fig. 3**a**. The remarkable surface-resolved spectroscopic consistency with our STM measurements in Fig. 2 confirms the validity of both experiment and calculation. We further find that the bulk states also feature this peak as shown in the lower panel of Fig. 3a, in strong support of the interpretation that the experimentally observed peak stems from the bulk band. Crucially, the calculation also provides momentum space insight into the origin of this peak, in that it arises from a nearly flat band which hybridizes with an electron-like band bottom due to SOC as shown in Fig. 3**b** and its inset. Consistent with previous bulk calculations identifying this material as a half-metallic ferromaget[17-20], the low energy states around the Fermi energy are all spin majority states (Fig. 3**b**).

We find that the essential momentum features of the flatband can be well captured by the fundamental kagome model (Fig. 3**c**). A tight-binding model for nearest neighbor hopping on the kagome lattice gives a kinetically frustrated flatband degenerate with a quadratically dispersing band at the origin of reciprocal space. The intrinsic breaking of time-reversal-symmetry splits the spin degenerate bands into two sets well separated in energy. SOC further lifts the band degeneracy at the $\Gamma$ point and makes originally flatband weakly dispersive. In this model, the flatband has a nontrivial Chern number carrying Berry phase effects[21-25].

As the peak on the S surface is the sharpest spectroscopic feature of the tunnelling signal and has origin in the kagome flatband, we further explore the spin-orbit related Berry phase effects by applying a strong magnetic field and performing spectroscopic measurements with high spatial and energy resolution. The *c* axis field dependent measurement at the same atomic site reveals that the peak shifts progressively to higher energies with increasing field magnitude independent of whether the field is parallel or antiparallel to the *c* axis (Figs. 4**a** and **b**). Fitting the field evolution of this shift with a linear function yields a slope of $0.17\pm0.01$meVT$^{-1}$ = $2.9\pm0.2\mu_B$ (Fig. 4**e**). Such an energy shift has been reproducibly observed in different samples. Moreover, the energy-shift $\Delta E_{8T-0T}$ over a clean S area is spatially homogeneous with an averaged value of 1.4meV (Figs. 4**c** and **d**), from which an averaged magnitude of the magnetic moment $3.0\pm0.3\mu_B$ can be derived from our data (Fig. 4**f**). The observed field induced shift of the flatband peak in this kagome magnet is unusual and unprecedented in the literature as we discuss below.

In a spin doubly-degenerate system, an applied magnetic field splits a density-of-state peak into two peaks separate in energy, as in the conventional Zeeman effect often seen in tunnelling experiments[27]. In the spin singly-degenerate case, the magnetic field will instead only shift the state in energy, the sign of which depends on the respective field orientation. When the magnetic moment of the state is polarized with an applied field in the presence of the ferromagnetic order, the state will always shift to the same direction in energy regardless of the relative field orientation (Fig. 4g upper panel, see SI for further elaboration). The systematic data trend we see (Figs. 4a-f) which is reproducible from sample to sample seems to suggest a magnetization-polarized scenario. If the magnetic moment has the same sign as the field, its energy should be lowered (Fig. 4g upper panel). Our measurements instead clearly show this peak shifting towards positive energies, indicating an effective negative magnetic moment of the state. Such a negative magnetic moment (-3μ$_B$) is highly nontrivial, as the conventional magnetization-polarized spin contribution is positive with a much smaller magnitude of ~1μ$_B$. This contrast indicates that there are orbital contributions in the magnetic response beyond the spin Zeeman effect, arising either from the atomic orbital angular momentum or the kinetic motion of the band electrons under SOC as discussed in Refs. 21-25. Considering the itinerant character of the Co 3$d$ electrons as a result of the short Co-Co bonding distance in this material[15,18], as well as the nontrivial Berry phase of the flatband, it is more likely that there exists a negative orbital magnetic moment resulting from the existence of SOC in the kagome flatband (lower panel in Fig. 4g).

Our first-principles calculation reveals the existence of orbital magnetism associated with the flatband with negative sign as shown in the upper panel of Fig. 4h (diamagnetic in this specific band, while the sample as a whole is ferromagnetic dominated by spin magnetization, and the net orbital magnetization of occupied bands is negligible). This negative sign is remarkably in agreement with our interpretation of the experimental results. In the magnetic kagome model, $m_{\text{orbital}}(\boldsymbol{k})$ for the flatband computed analytically within a $\boldsymbol{k}\cdot\boldsymbol{p}$ approximation is found to be proportional to $-\frac{k^2 t^2 \lambda}{k^4 t^2 + 48\lambda^2}$ (for both spin-majority band polarization directions along those of the applied magnetic field) as shown in the lower panel of Fig. 4h, where λ is the SOC strength and $t$ is the nearest-neighbor hopping integral. This offers a heuristic reference for understanding the negative orbital magnetism tied to the sign of the microscopic SOC, which also determines the sign of the Berry curvature. Since the SOC term is proportional to $L\cdot S$, this orbital magnetic moment is also tied to the spin magnetization polarized by the applied magnetic field. Thus the associated field driven energy shift follows the magnetization-polarized scenario depicted in Fig. 4g. As there can also be electron-electron interaction for 3$d$ orbitals involving multiple bands, a complete and quantitative understanding of this emergent behavior requires a comprehensive quantum many-body theory of correlated electrons on the kagome lattice with strong magnetic polarization and SOC. Crucially, the orbital magnetism of the flatband is a generic property of the kagome lattice in the ferromagnetic state with SOC, and can therefore be generally applied to a large family of transition metal based kagome materials, whose electronic structures are of recent interest[6,28-30]. Our clear observation of a sharp peak located just below the Fermi level with strong Berry phase effects also indicates that the nearby flatband can contribute substantially to the anomalous low-energy physics, and therefore be probed readily in transport. Identification of Berry curvature effect as such is also crucial for exploring giant linear and nonlinear optical effects in future experiments. Furthermore, this new method of characterizing the orbital magnetism in a band or energy resolved fashion we demonstrate here can be used as a more direct probe of Berry phase effects in most bulk quantum materials that are of current interst[1-7].

Our experiments taken together demonstrate an unusual field response of the electronic excitations in the half-metallic kagome magnet, uncovering hitherto unknown quantum phenomena including a large negative orbital magnetic moment of the flatband originating from the intrinsic Berry curvature effects. While the fundamental relationship between orbital magnetism and Berry phase effects has been theoretically studied for decades[21-25], the decisive experimental evidence at the microscopic level in a band-resolved way has long been lacking which is essential to explore their band topology. Here, our atomically-layer-resolved and magnetic-field-dependent spectroscopic method not only demonstrates this remarkable connection directly at the experimental level but also paves pathways for future band-resolved and field-perturbed explorations on topological magnets and superconductors where Berry phase and emergent phenomena are intertwined. An experimental handle of these effects in exploring quantum materials in general is a crucial milestone in the field of strong correlation and quantum topology.

**Figures and captions**

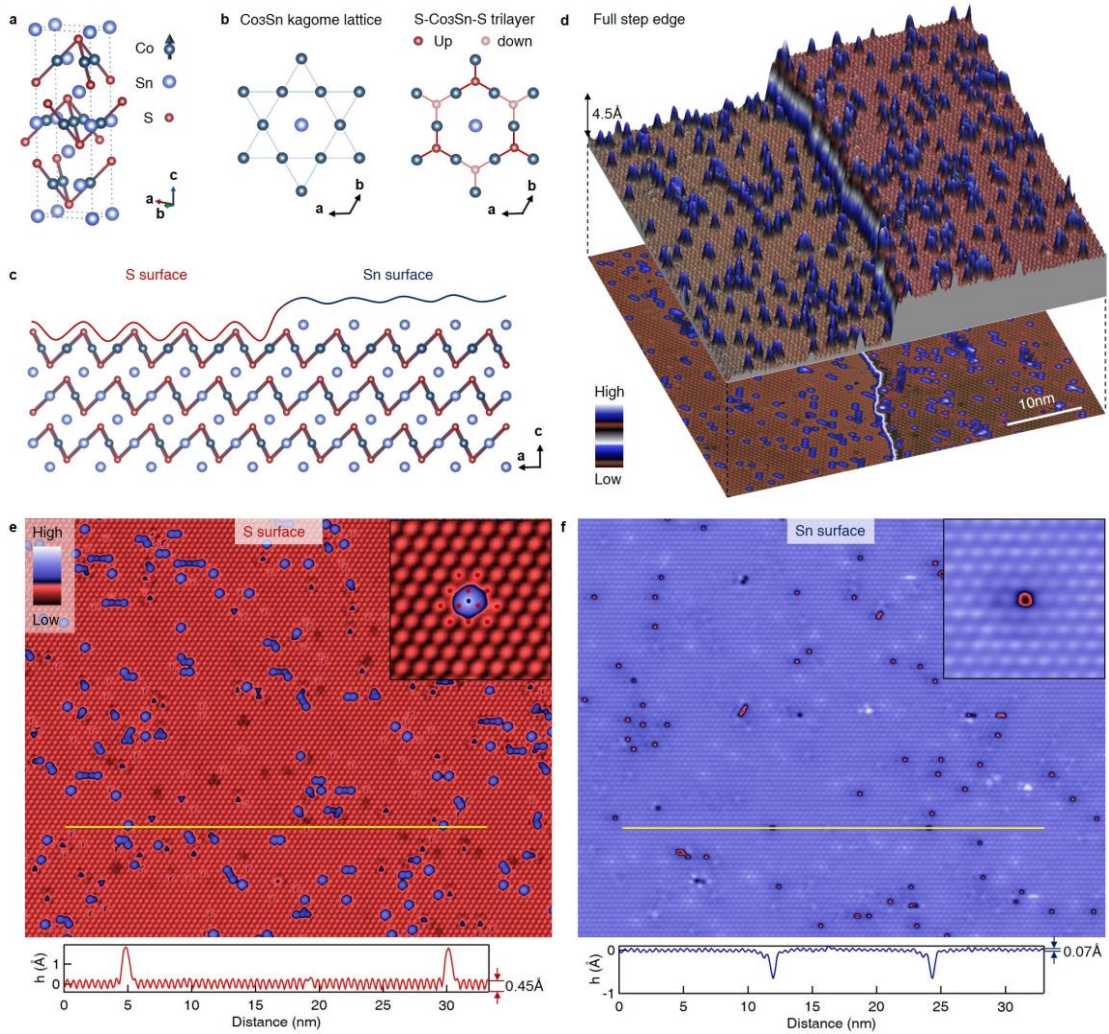

**Figure.1 | Atomic scale visualization of the cleavage surface of Co₃Sn₂S₂. a,** Crystal structure of $Co_3Sn_2S_2$. The crystal has a ferromagnetic ground state with spins on Co atoms aligned along the *c* axis. **b,** Kagome lattice structure of the $Co_3Sn$ layer (left panel) and its bonding with the adjacent S layers (right panel). **c,** Side view of the crystal structure and illustration of the two possible terminating surfaces. The curve illustrates the surface profile. **d,** Topographic image of a full step between two surfaces of the same kind and its three-dimensional illustration. The step edge height is ~4.5Å, consistent with *c*/3. **e,** Topographic image of the S surface. The inset shows the common defect (Sn adatom) on this surface, which sits at the hexagonal-close-packed site. The lower panel shows the line profile as marked on the S surface. **f,** Topographic image of the Sn surface. The inset shows the common defect (Sn vacancy) on this surface. The lower panel shows the line profile as marked on the Sn surface.

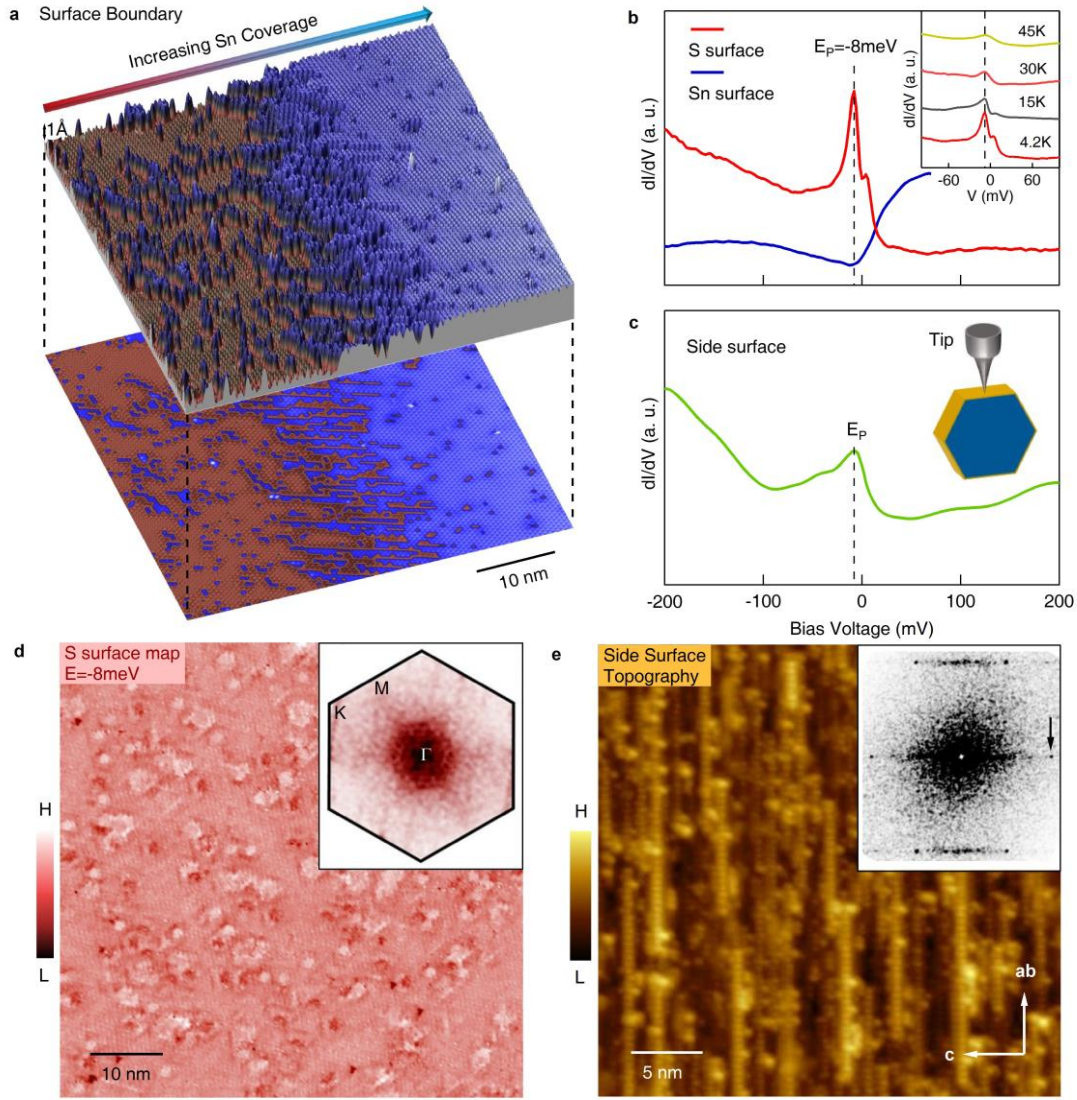

**Figure. 2 | Observation of a sharp peak around the Fermi energy. a,** Surface boundary between the Sn surface and S surface and the three-dimensional view. The S surface smoothly evolves to become the Sn surface with the increased coverage of Sn adatoms. **b,** Differential conductance spectrum taken on S (red) and Sn (blue) surfaces away from any defects. Inset: temperature evolution of the spectrum taken on the S surface. **c,** Averaged spectrum taken on the side cleaving surface. Inset: illustration of the tunnelling geometry. **d,** Differential conductance map on the S surface at E = -8meV. Inset: Fourier transform image. **e,** Topographic image of the side surface showing a stripe like atomic structure. Inset: Fourier transform image, where the arrow marks the Bragg peak corresponding to $(c/3)^{-1}$.

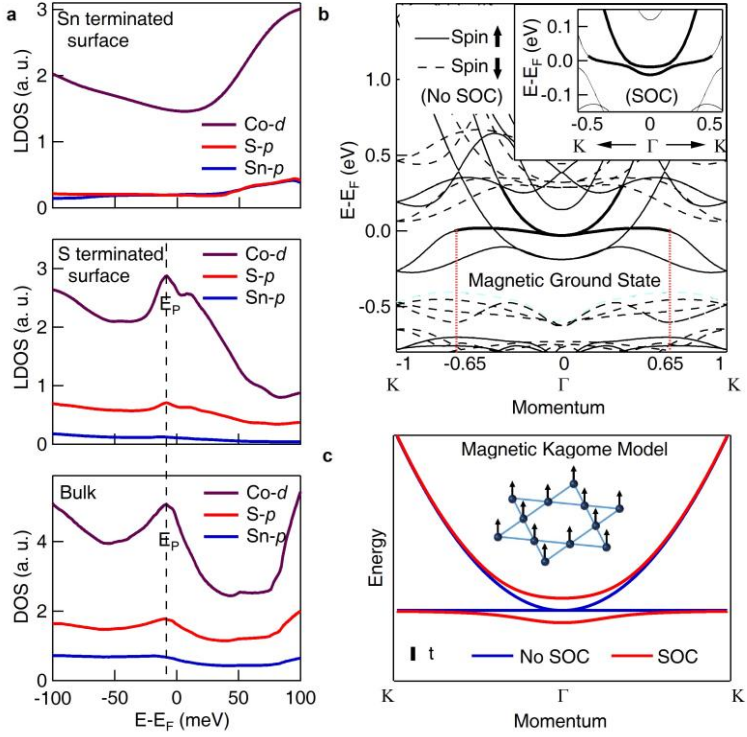

**Figure. 3 | Magnetic kagome flatband nature of the density-of-state peak. a,** Calculated LDOS on the Co (within the $Co_3Sn$ layer), S (within the S layer) and Sn (within the Sn layer) atoms at the Sn-terminated surface (upper panel) and S-terminated surface (middle panel), respectively. Lower panel: calculated DOS on the Co, S and Sn atoms in the bulk. All the above calculations include SOC. **b,** Calculated spin-resolved band structure without SOC at $k_Z=0$, showing a nearly flat band degenerate with an electron-like band bottom near the Fermi level. With SOC included, there is a further hybridization between these two bands (inset), and their accumulated states contribute to the LDOS peak at $E_P$. **c,** Tight-binding model of single layer magnetic kagome with (red) and without (blue) SOC (coupling strength $\lambda = 0.2t$ where $t$ is nearest-neighbor hopping integral). The inset illustrates the kagome lattice with ferromagnetism.

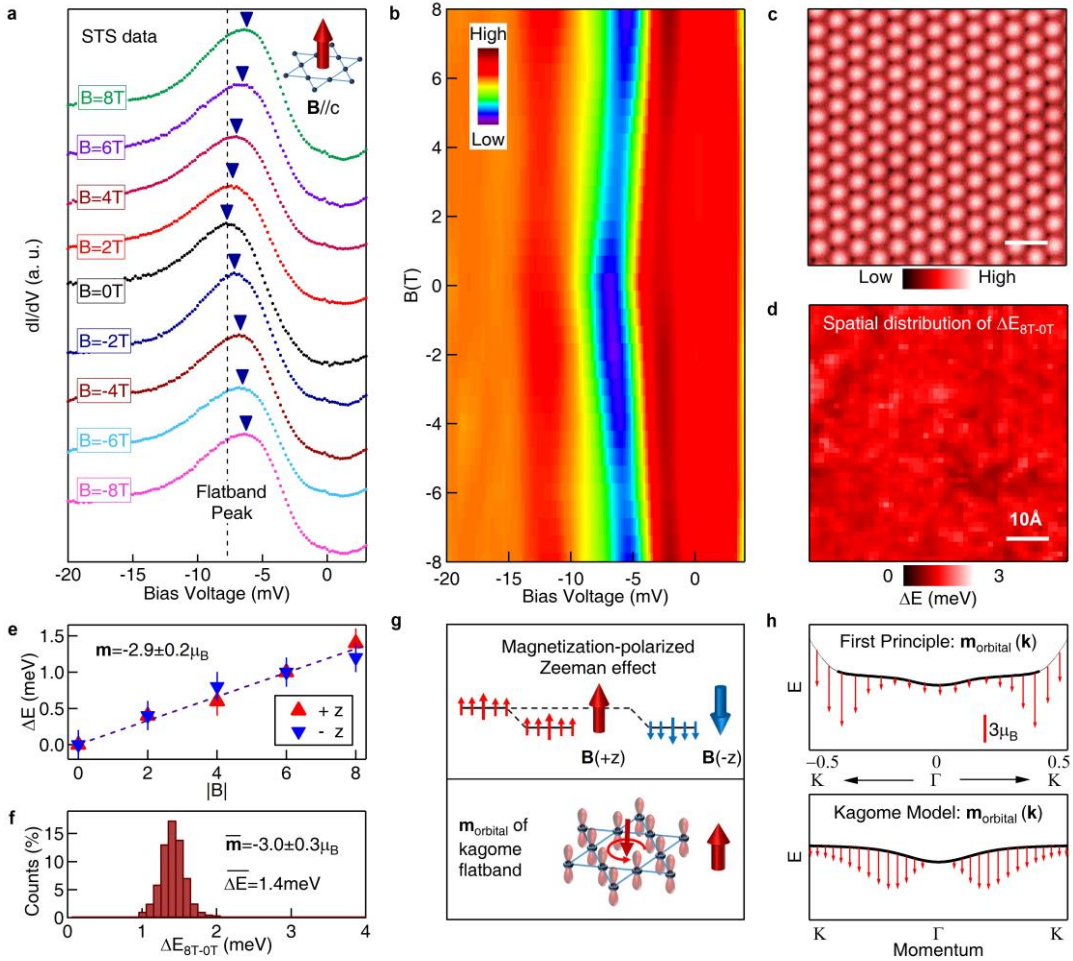

**Figure. 4 | Negative orbital magnetism of the flatband peak. a,** Magnetic field dependence of the peak on S surface. The peak exhibits a shift in energy of equal magnitude with the field applied in either *c* axis direction. The inset illustrates the magnetic field perpendicular to the kagome lattice. **b,** Intensity plot of second derivate of interpolated field-dependent tunneling spectra, showing similar shift as **a**. **c,** Topographic image of a clean S surface. **d,** Energy-shift map between 8T and 0T for this surface. **e,** Energy-shift of peak as a function of external field, from which the magnetic moment value can be derived as $m = -2.9 \pm 0.2\mu_B$. Error bars denote energy resolution. **f,** Histogram of the energy-shift over the area in **c** and **d**, showing an average value of 1.4meV that gives $m = -3.0 \pm 0.3\mu_B$. **g,** Upper panel: illustration of the magnetization-polarized Zeeman effect. The applied field aligns the spins in the same orientation and the energy shift is the same for field applied in both orientations (see SI for further elaboration). Lower panel: illustration of the large negative orbital magnetism of the flatband in the kagome lattice. **h,** Upper panel: orbital magnetism for the flatband calculated from first principles (full band structure shown in Fig. 3**b**). The magnetic moment (red arrows) is plotted along the flatband. The red bar marks the units of the magnetic moment value. Lower panel: orbital magnetism from the magnetic kagome lattice model. The magnetic moment (red arrows, arbitrary unit) is plotted along the flatband.

22. Xiao, D., Chang, M. C. and Niu, Q. Berry phase effects on electronic properties. *Rev. Mod. Phys.* **82**, 1959-2007 (2010).
23. Resta, R. Electrical polarization and orbital magnetization: the modern theories. J*ournal of Physics: Condensed Matter* **22**, 123201 (2010).
24. Thonhauser, T. Theory of Orbital Magnetization in Solids. *Int. J. Mod. Phys. B* **25**, 1429 (2011).
25. Vanderbilt, David. Berry Phases in Electronic Structure Theory: Electric Polarization, Orbital Magnetization and Topological Insulators. (Cambridge University Press, 2018).
26. Fischer, Ø. *et al.* Scanning tunneling spectroscopy of high-temperature superconductors. *Rev. Mod. Phys.* **79**, 353 (2007).
27. Ternes, M., Heinrich, Andreas. J. & Schneider, W. -D. Spectroscopic manifestations of the Kondo effect on single adatoms. *J. Phys.: Condens. Matter* **21**, 053001 (2009).
28. Lin, Z. *et al.* Flatbands and Emergent Ferromagnetic Ordering in $Fe_3Sn_2$ Kagome Lattices. *Phys. Rev. Lett.* **121**, 096401 (2018).
29. Nayak, A. K. *et al.* Large anomalous Hall effect driven by a nonvanishing Berry curvature in the noncolinear antiferromagnet $Mn_3Ge$. *Sci. Adv.* **2**, e1501870 (2016).
30. Kuroda, K. *et al.* Evidence for magnetic Weyl fermions in a correlated metal. *Nat. Mater.* 16, 1090–1095 (2017).
**Acknowledgements** Experimental and theoretical work at Princeton University was supported by the Gordon and Betty Moore Foundation (GBMF4547/ Hasan) and the United States Department of energy (US DOE) under the Basic Energy Sciences programme (grant number DOE/BES DE-FG-02-05ER46200). Work at Renmin University was supported by the Ministry of Science and Technology of China (2016YFA0300504), the National Natural Science Foundation of China (No. 11574394, 11774423, 11822412), the Fundamental Research Funds for the Central Universities, and the Research Funds of Renmin University of China (RUC) (15XNLF06, 15XNLQ07, 18XNLG14). S.T. and T.N. acknowledge supports from the European Union's Horizon 2020 research and innovation program (ERC-StG-Neupert-757867-PARATOP). Z.W and K.J. acknowledge US DOE grant DE-FG02-99ER45747. We also acknowledge the Natural Science Foundation of China (grant numbers 11790313 and 11774007), the Key Research Program of the Chinese Academy of Sciences (grant number XDPB08-1), National Key R&D Program of China (grant numbers 2016YFA0300403 and 2018YFA035601), Princeton Center for Theoretical Science (PCTS) and Princeton Institute for the Science and Technology of Materials (PRISM)'s Imaging and Analysis Center at Princeton University. Muon Spin Rotation (μSR) experiments were performed at the πE3 beamline of the Paul Scherrer Institute, using the HAL-9500 μSR spectrometer. Z.G. acknowledges Robert Scheuermann for the support in μSR experiments. M.Z.H. acknowledges support from Lawrence Berkeley National Laboratory and the Miller Institute of Basic Research in Science at the University of California, Berkeley in the form of a Visiting Miller Professorship.
**Author contributions** J-X.Y. and S.S.Z. conducted the STM and STS experiments in consultation with M.Z.H.; Q.W., H.L., H.Z. and S.J. synthesized and characterized the sequence